\documentclass[a4paper, 10pt]{article}

\usepackage{epsfig}
\usepackage{graphicx}
\usepackage{amsmath}
\usepackage{amssymb}
\usepackage{mathrsfs}

\newtheorem{theorem}{Theorem}[section]

\newtheorem{definition}{Definition}[section]
\newtheorem{claim}{Claim}[section]
\newtheorem{proposition}{Proposition}[section]

\newcommand{\proofends}{\vspace{-0.1in}\begin{flushright} \mbox{}\hfill{$\square$} \end{flushright}}
\newcommand{\beginproof}{\vspace{0.1in}\textbf{Proof:}\\}

\begin{document}

\title{Multi-Armed Bandit Mechanisms for Multi-Slot \\Sponsored Search Auctions}

\author{Akash Das Sarma\\Indian Institute of Technology,\\Kanpur, 208016.
\and
Sujit Gujar\\
Indian Institute of Science,\\ Bangalore, 560012.
\and
Y Narahari\\
Indian Institute of Science,\\ Bangalore, 560012.
}

\maketitle

\begin{abstract}
In pay-per click sponsored search auctions which are currently
extensively used by search engines, the auction for a keyword
involves a certain number of advertisers (say $k$) competing for
available slots (say $m$) to display their ads. This auction
is typically conducted for a number of rounds (say $T$).
There are click probabilities $\mu_{ij}$ associated with
each agent-slot pairs. The goal of the search engine is to maximize
social welfare of the advertisers, that is, the sum of values of the
advertisers. The search engine does not know the true
values advertisers have for a click to their respective  ads
and also does not know the
click probabilities $\mu_{ij}$s.
A key problem for the search  engine therefore is to
learn these click probabilities during the $T$ rounds of
the auction and also to ensure that the auction mechanism is
truthful. Mechanisms for addressing such learning and incentives 
issues have recently been introduced and are aptly referred to as
multi-armed-bandit (MAB) mechanisms. When  $m=1$,
characterizations for truthful MAB mechanisms are available in 
the literature and it has been shown that the regret
for such mechanisms will be $O(T^{\frac{2}{3}})$.
In this paper, we seek to derive a
characterization in the realistic but non-trivial
general case when $m>1$ and obtain
several interesting results. Our contributions include:
(1) When $\mu_{ij}$s are {\em unconstrained\/},
we prove that any truthful mechanism must satisfy
{\em strong pointwise monotonicity\/} and show that 
the regret will be $O(T)$ for such mechanisms.
(2) When the clicks on the ads follow a certain {\em click precedence
property\/}, we show that {\em weak pointwise monotonicity\/} is necessary for
MAB mechanisms to be truthful.
(3) If the search engine has a certain coarse pre-estimate of $\mu_{ij}$
values and wishes to update them during the course of the $T$ rounds,
we show that {\em weak pointwise monotonicity\/} and  {\em weakly separatedness\/}
are necessary and sufficient conditions for the MAB mechanisms
to be truthful.
(4) If the click probabilities are separable into
agent specific and slot specific terms,
we provide a
characterization of MAB mechanisms that are {\em truthful in expectation\/}.
\end{abstract}




\section{Introduction}
\label{sec:Introduction}
Whenever a user searches any set of keywords on a search engine,
along with the search results, called \emph{organic results},
the search engine displays advertisements related to those
keywords on the right side of the organic results.
In pay-per-click sponsored search auctions, the
search engine charges an advertiser for displaying her ad only if a
user clicks on her ad.
The decision regarding which ads are be displayed
and their respective order is based on the bids submitted by the advertisers
indicating the maximum amount they are willing to pay per click.
To perform any optimizations, such as maximizing social welfare
or maximizing revenue to the search engine,
the true valuations of the advertisers are needed.
Being rational, the advertisers may actually
manipulate their bids and therefore a primary goal of
the search engine is to design an auction for which
it is in the best interest of each advertiser to bid truthfully
irrespective of the bids of the other advertisers.
Such an auction is said to be \emph{Dominant Strategy Incentive
Compatible} (DSIC), or truthful.

These auctions also take into account crucially the click probabilities
or clickthrough rates (CTRs). Given an agent $i$ and a slot $j$, the click
probability $\mu_{ij}$ is the probability with
which the ad of agent $i$ will be clicked if the ad appears in slot $j$.
If the search engine knows the CTRs, then its problem is only to
design a truthful auction.
However, the search engine may not know the CTRs beforehand.
Thus the problem of the search engine is two fold: (1) learn the CTR values
(2) design a truthful auction.
Typically, the same set of agents compete for the given set of keywords.
The search engine can exploit this fact to learn the CTRs
by initially displaying ads by various advertisers. 
Also note, it is reasonable to assume that they may not revise their bids frequently.If the advertisers
were bidding true values, the search engine's problem would
have been the same as that of a multi-armed bandit (MAB)
problem \cite{ROBBINS52} for learning the CTRs. Since the agents may not report their true values,
the problem of the search engine can be described as one of designing an
incentive compatible MAB mechanism.
In the initial rounds, the search engine displays
advertisements from all the agents to learn the CTRs.
This phase is referred to as \textit{exploration} phase. Then
it uses the information gained in these rounds to maximize the
social welfare. The latter phase is referred to as
\textit{exploitation}. The search engine will invariably lose a part of
social welfare for the exploration phase. The difference between
the social welfare the search engine would have achieved with
 the knowledge of CTRs and the actual social welfare achieved
by a MAB mechanism is referred to as \emph{regret}. Thus,
regret analysis is also important while designing a
MAB mechanism.

\subsection{Related Work}
The problems where the decision maker has to optimize 
his total reward based on gained information as well as gain knowledge 
about the available rewards are referred to as Multi-Armed Bandit (MAB) 
problem. The MAB problem was first studied by 
Robbins \cite{ROBBINS52} in 1952. After his seminal work,
MAB problems have been extensively studied for regret analysis
and convergence rates. Readers are referred to \cite{AUER02}
for regret analysis in finite time MAB problems. However,
when a mechanism designer has to consider strategic behavior
of the agents, these bounds on regret would not work.
Recently,
Babaioff, Sharma, and Slivkins
\cite{BABAIOFF09} have derived a characterization for truthful
MAB mechanisms in the context of pay-per-click sponsored search auctions
if there is only a single slot for each keyword.
They have shown that any truthful MAB mechanism must have at least
$\Omega(T^{\frac{2}{3}})$ worst case regret and also proposed a mechanism
that achieves this regret. Here $T$ indicates the number of rounds
for which the auction is conducted for a given keyword, with the same set of agents
involved.

Devanur and Kakade \cite{DEVABUR09a} have also addressed the
problem of designing truthful MAB
mechanisms for pay-per-click auctions with a single sponsored slot.
Though they have
not explicitly attempted a characterization of
truthful MAB mechanisms,
they have derived similar results
on payments as in \cite{BABAIOFF09}. They
have also obtained a bound on regret of a
MAB mechanism to be  $O(T^{\frac{2}{3}})$. Note that
the regret in \cite{DEVABUR09a} is regret in the revenue
to the search engine, as against regret analysis in \cite{BABAIOFF09}
is for social welfare of the advertisers. In this paper, 
unless explicitly stated, when we refer to \textit{regret}, 
we mean loss in social welfare as compared to social welfare that could 
have been obtained with known CTRs.

In both of the above
papers, only a single slot for advertisements is considered
and therefore the practical appeal is limited.
Generalization of their work to the more realistic case of multiple
sponsored slots is non-trivial and our paper seeks to 
fill this research gap.

Prior to the above two papers, Gonen and Pavlov \cite{GONEN07}
had addressed the issue of unknown CTRs in multiple slot
sponsored search auctions and proposed a specific mechanism.
Their claim that their mechanism is truthful in expectation
has been contested by \cite{BABAIOFF09,DEVABUR09a}.
Also Gonen and Pavlov do not provide any
characterization for truthful multi-slot MAB mechanisms.

\subsection{Our Contributions}
In this paper, we extend the results of
Babaioff, Sharma, and Slivkins \cite{BABAIOFF09}
and Devanur and Kakade \cite{DEVABUR09a}
to the non-trivial general case of two or more sponsored slots.
The precise question we address is:
{\em which MAB mechanisms for multi-slot
pay-per-click sponsored search auctions are
dominant strategy incentive compatible (or truthful)?\/} We describe
our specific contributions below.

In the first and most general setting (Section \ref{ssec:general}),
we  assume no knowledge of click through rate ($\mu_{ij}$) values or any
relationships among $\mu_{ij}$ values. We refer to this
setting as the ``unknown and unconstrained CTR'' setting.
Here we show that any truthful mechanism must satisfy a highly
restrictive property which we refer to as {\em strong pointwise
monotonicity\/} property. We show that all
mechanisms satisfying this property will however
exhibit a high regret, which is $O(T)$. This immediately
motivates our remaining Sections
\ref{ssec:general2}, \ref{ssec:agentdependantctr}, and
\ref{ssec:separablectr}, where we explore the following
variants of the
general setting which yield more reasonable characterizations.

First, in Section \ref{ssec:general2}, we consider a setting where the realization
is restricted according to a property which we call the \textit{Higher Slot Click Precedence}
property (a click in a lower slot will automatically imply that a
click is received if the same ad is shown in any higher slot). For this
setting, we provide a weaker necessary condition than strong pointwise monotonicity. Finding a necessary and sufficient condition however remains open.

In Section \ref{ssec:agentdependantctr}, we provide a complete characterization of
MAB mechanisms which are {\em truthful in expectation\/} under
a stochastic setting where a coarse estimate of $\mu_{ij}$
is known to the auctioneer and to the agent $i$, perhaps from
some database of past auctions. Under this setting, the auctioneer
updates his database of $\mu_{ij}$ values based on the observed
clicks, thereby improving his estimate and maximizing revenue.

Finally, in Section \ref{ssec:separablectr}, we derive a complete
characterization of truthful multi-slot MAB mechanisms for
a stochastic setting where we assume that the $\mu_{ij}$s are
separable into agent-dependent and slot-dependent parts.
Here, unlike the previous setting, we do not
assume existence of any information on agent-dependent click
probabilities.

For all the above multi-slot sponsored search auction
settings, we show that the slot allocation in truthful mechanisms
must satisfy some notion of monotonicity with respect to the
agents' bids and a certain weak separation between exploration
and exploitation.

Our results are summarized in Table \ref{tab:results}.
\begin{table*}[htb]
\centering
\begin{tabular}{|c|c|c|l|c|}
\hline
Number of & Learning Parameter & Solution Concept & Allocation rule & Regret \\
Slots ($m$) & (CTR) & & & \\
\hline
$m = 1$ \cite{BABAIOFF09} & Unrestricted & DSIC & Pointwise monotone and &  O($T^{2/3}$) \\
 & & & Exploration separated & \\
\hline
$m>1$ & Unrestricted & DSIC & Strongly pointwise monotone & O(T) \\
 & & &  and weakly separated & \\
 & & & & \\
 & Higher Slot Click Precedence & DSIC & Weakly pointwise monotone & 
					regret analysis  \\
 & & & and weakly separated & not carried out\\
 & & & (Necessary Condition) & \\
 & & & & \\
 & CTR Pre-estimates available & Truthful in & Weakly Pointwise monotone & regret analysis \\
 & & expectation&  and weakly separated & not carried out\\
 & & & & \\
 & Separable CTR & Truthful in  & Weakly Pointwise monotone& 
  O($T^{2/3}$)\\
 & & expectation &  and weakly separated & (Experimental Evidence)\\
\hline
\end{tabular}
\caption{Results}
\label{tab:results}
\end{table*}

Our approach and line of attack in this paper follow that of \cite{BABAIOFF09} where
the authors use the
notions of \emph{pointwise monotonicity}, \emph{weakly
separatedness} and {\it exploration separatedness} quite critically
in characterizing truthfulness. Since our paper deals
with the general problem of which theirs is a special case,
these notions continue to play an important role in our paper.
However, there are some notable differences as explained below.
We generalize their notion of {\it pointwise
monotonicity} in two ways. 
The first notion we refer to as strong pointwise monotonicity and the second one 
as weak pointwise monotonicity. In addition to this,
we introduce the key notions of {\it Influential Set},
$i${\it -influentiality} and {\it Strongly influential}.
We  use these new notions to define a non-trivial
generalization of their notion of {\it weakly separatedness},
to which we, however, continue to associate the same name.
The characterization of truthful mechanisms for a single
parameter was provided  by \cite{ARCHER01,MYERSON81}.
For deriving payments to be assigned to the agents for
truthful implementation, we use the approach in
\cite{ARCHER01,MYERSON81}.

In Section \ref{sec:experiments}, we provide some
simple experimental results on regret analysis.
We conclude the paper in Section \ref{sec:con}.

\section{System Set up and Notation}
\label{sec:def}
In the auction considered, there are $k$ agents and $m$ ad
slots ($k\geq m$). Each agent has a single advertisement that
she wants to display and a private value $v_i$ which is her
value per click on the ad. The auctioneer, that is the search
engine wishes to distribute the ads among these slots. These
advertisements have certain click probabilities which depend upon
the agent as well as the slot with which the agent is associated.
Let $\mu_{ij}$ be the probability of an  ad of an agent $i$
receiving click in slot $j$. Now, the goal of the search engine
is to assign these agents to the slots in such a way that the
social welfare, which is the total value received by the
bidders, is maximized. However, there are two problems,
(i) the search engine does not know $v_i$, the valuations
of the agents and (ii) the search engine may not know the
click probabilities $\mu_{ij}$.

So, the goal of the search engine is:
(i) to design a DSIC auction in which it is in the agents'
interest to bid their true values, $v_i$'s
(ii) to estimate $\mu_{ij}$.
We consider multi-round auctions, where the search
engine displays the various advertisements repeatedly over a large
number of rounds. The mechanism uses the initial rounds in an
{\it explorative} fashion to learn $\mu_{ij}$ and then uses the
other rounds {\it exploitatively} to gain value.

The system works as follows. At the start of the auction,
each agent submits a sealed bid $b_i$. Based on this bid
and the click information from previous rounds, the mechanism
decides to allocate each ad slot to a particular agent and
then displays the $m$ chosen ads. The user can now click on
any number of these ads and this information gets registered
by the mechanism for future rounds. At the end of $T$ rounds,
depending on the bids submitted by the agents and the number
of clicks received by each agent, the agents have to make a
certain payment $P_i$ to the mechanism.

\noindent\textit{Note: $P_i$ and $C_i$ are functions of $b$
and $\rho$. Whenever the arguments are clear from the context,
we just refer to them as $P_i$ and $C_i$.}

A mechanism can be formally defined as the tuple $(A,P)$
where $A$ is the allocation rule specifying the slot allocation
and $P$ is the payment rule.

The important notation used in the paper is summarized in
Table \ref{tab:notation}. Following this, we define the terms used in
this paper.

\begin{table}[t]
{\small
\centering
\begin{tabular}{|c|l|}
\hline
$K$ & = $\{1,2,\ldots,k\}$, Set of agents \\
\hline
$M$ & = $\{1,2,\ldots,m\}$ Set of slots\\
\hline
$i$ & Index of an agent, $i=1,2,\ldots,k$\\
\hline
$j$ & Index of a slot, $j=1,2,\ldots,m$\\
\hline
$T$ & Total number of rounds\\
\hline
$t$ & A particular round. $t \in \{1,2,\ldots,T\}$\\
\hline
$A_{ij}(t)$ &= 1 If an agent $i$ is allocated slot $j$ in round $t$
\\ &= 0 otherwise \\
$A(t)$ & $(A_{ij}(t))_{i\in K,j\in M}$\\
\hline
$A$ & $=(A(1),A(2),\ldots,A(T))$, Allocation rule\\
\hline
 $\rho_{ij}(t)$ & = 1 if agent $i$ gets a click in slot $j$
 in round $t$\\ & = 0 otherwise \\ $\rho(t)$ &
$(\rho_{ij}(t))_{i\in K, j \in M}$\\
\hline
$\rho$ & = $(\rho(1), \rho(2), \ldots, \rho(T))$\\
\hline
$v_i$ & Agent $i$'s valuation of a click to her ad\\
\hline
$b_i$ & Bid by agent $i$\\
\hline
$b$ & Bid vector, indicating bids of all the agents\\
 & $=(b_i,b_{-i})=(b_1,b_2,\ldots,b_k)$\\
\hline
$C_i(b,\rho)$ & Total number of clicks obtained by an agent $i$\\
 & in $T$ rounds\\
\hline
$P_i(b,\rho)$ & Payment made by agent $i$\\
\hline
$P(b,\rho)$ & $=(P_1(\ldotp),P_2(\ldotp),\ldots,P_k(\ldotp))$,
 Payment rule\\
\hline
$U_i(v_i,b,\rho)$ & Utility of an agent $i$ in $T$ rounds \\
 & = $v_iC_i(b,\rho) - P_i(b,\rho)$\\
\hline
$b_i^+$ & A real number > $b_i$\\
\hline
$\alpha_i$ & Click probability associated with agent $i$\\
\hline
$\beta_j$ & Click probability associated with slot $j$\\
\hline
$\mu_{ij}$ & The probability that an ad of an agent $i$ receives\\
 & click when the agent is allotted slot $j$.\\
\hline
$N(b,\rho,i,t)$ & Set of slot agent pairs in round $t$\\
 & that influence agent $i$ in some future rounds\\
\hline
CTR & Click Through Rate (Click Probability)\\
\hline
DSIC & Dominant Strategy Incentive Compatible\\
\hline
\end{tabular}
\caption{Notation} \label{tab:notation}
}
\end{table}

\subsection{Important Notions and Definitions}
\label{def}
\begin{definition}
[Realization $\rho$]We define
a realization $\rho$ as a vector $(\rho(1),\rho(2),\ldots,
\rho(T))$ where $\rho(t) = [\rho_{ij}(t)]_{K\times M}$
is click information in round $t$. $\rho_{ij}(t) =1$,
if an agent $i$'s ad receives a click in slot $j$ in round $t$,
else $0$.
\end{definition}
It is to be noted that the mechanism observes only those
$\rho_{ij}(t)$ where $A_{ij}({b},\rho,t)=1$.

\begin{definition}
[Clickwise Monotonicity] We \\call an allocation rule $A$
clickwise monotone if for a fixed $(b_{-i},\rho)$, the number
of clicks, $C_i(b_i,b_{-i},\rho)$ is a non-decreasing function
of $b_i$. That is,  $\frac{dC_i(\ldotp)}{db_i}\geq 0 \; \forall \: (b_{-i},\rho)$.
\end{definition}

\begin{definition}[Weak Pointwise Monotonicity]
We call an allocation rule weak pointwise
monotone if, for any given $(b_{-i},\rho)$, and bid $b_i^+>b_i$,
$A_{ij}((b_i,b_{-i}),\rho,t) = 1 \Rightarrow$\\ $
A_{ij'}((b_i^+,b_{-i}),\rho,t) = 1$ for some slot $j'\leq
j$, $\forall t$.
\end{definition}

\begin{definition}
[Influential Set] Given a bid vector, $b$,
a realization $\rho$ and round $t$, an influential
set $I(b,\rho,t)$ is the set of all agent-slot allocation pairs
$(i,j)$, such that (i) $A_{ij}({b},\rho,t)=1$ and (ii) a change
in $\rho_{ij}(t)$ will result in a change in the allocation
in a future round.  $t$ is referred to as an \emph{influential
round}. Agent $i$ is referred to as an \emph{influential agent} and
$j$ as \emph{influential slot} w.r.t round $t$.
\end{definition}

\begin{definition}
[$i$-Influential Set]
We define the $i$-\emph{influential set}
$N(b,\rho,i,t)\subseteq I(b,\rho,t)$ as the set of all influential
agent-slot pairs $(i',j')$ such that change in $\rho_{i'j'}(t)$
will change the allocation of agent $i$ in some future round.
\end{definition}

\begin{definition}
[Strongly Influential] We call a slot-agent pair $(i^*,j^*)$
\emph{strongly influential} in round $t$ w.r.t. the realization
$\rho(t)$, if changing the realization (toggling) in the bit
$\rho_{i^*j^*}(t)$ changes the allocation in a future round.
We call such a set $(i^*,j^*,t)$ \emph{strongly $i$-influential}
if one of its influenced agents is $i$.
\end{definition}

\begin{definition}
[Weakly Separated] We call an allocation rule weakly separated
if for a given $(b_{-i},\rho)$ and two bids of agent $i$, $b_i$ and
$b_i^+$ where $b_i<b_i^+$, $N((b_i,b_{-i}),\rho,i,t) \subseteq
N((b_i^+,b_{-i}),\rho,i,t)$.
\end{definition}
This means that when an agent $i$ increases her bid, while the
other parameters are kept fixed, the allocation in
the originally influential slots does not change, only new
influential agent-slot pairs can get added. We continue to
use definitions of \emph{Normalized Mechanism}
and \emph{Non-degeneracy} from \cite{BABAIOFF09}.
With these preliminaries, we are now ready to characterize truthful
MAB mechanisms for various settings in the next section.

\section{Characterization of Truthful MAB Mechanisms}
\label{sec:char}
Before stating our results, we prove a minor claim that we will
use to develop our characterizations. We will use this claim
implicitly in our proofs.

\begin{claim}
\label{clm:str_inf}
Given $(b,(\rho(1),\rho(2),\ldots,\rho(t-1)))$,
if $(i^*,j^*)$ is \emph{$i$-influential} in round $t$, then
$\exists \rho^*(t)$ such that $(i^*,j^*)$ is also
{\it strongly $i$-influential} w.r.t. $\rho^*(t)$ in round $t$.
\end{claim}
\beginproof
Suppose the claim is false. Let the
$i$-influential set of slots in round $t$ be $N({\it
b},\rho,i,t)=\{(i^1,j^1),(i^2,j^2),\ldots,\\(i^l,j^l),(i^*,j^*)\}$.
$N({b},\rho,i,t)\neq \phi$ since it has at least one element
$(i^*,j^*)$. Since we have assumed our claim to be false,
$(i^*,j^*)$ is not strongly $i$-influential for {\it any }
realization
$(\rho_{i^1j^1}(t), \rho_{i^2j^2}(t), \ldots, \rho_{i^lj^l}(t))$
or the allocation of agent $i$ in future rounds is the same
whether $\rho_{i^*j^*}$ is $0$ or $1$ for every given
$(\rho_{i^1j^1}(t),\rho_{i^2j^2}(t),\ldots,\rho_{i^lj^l}(t))$.
This means that the allocation of agent $i$ is the same in future
rounds for all realizations
$(\rho_{i^1j^1}(t), \rho_{i^2j^2}(t), \ldots,\rho_{i^lj^l}(t),
\rho_{i^*j^*}(t))$. But this contradicts the fact that
$\{(i^1,j^1), (i^2,j^2), \ldots,(i^l,j^l), (i^*,j^*)\}$ is
the set of $i$-influential slot-agent pairs in round $t$.
This proves our claim. 
\proofends
In our characterization of truthfulness under various
settings, we show that a truthful allocation rule $A$ must be
weakly separated. Though the proofs look similar, there are
subtle differences in each of the following subsections.
In our proofs, we start with the assumption that a truthful
allocation rule $A$ is not weakly separated. That is, \\
\begin{equation}
\label{eqn:contr_weak_sep}
\begin{array}{|l|}
    \hline
    \exists b_i<b_i^+, b_{-i},\rho,t \: \ni
    N(b_i,b_{-i};\rho,t,i)\not\subseteq N(b_i^+,b_{-i};\rho,t,i)\\
    \Rightarrow \exists (i^*,j^*) \in N(b_i,b_{-i};\rho,t,i) \ni \;
    (i^*,j^*)\not\in N(b_i^+,b_{-i};\rho,t)
    \\ \hline
\end{array}
\end{equation}
Subsequently, we show that this leads to a contradiction in each of the
subsections, implying the necessity of weakly separatedness.

\subsection{Unknown and Unconstrained CTRs}
\label{ssec:general}
In this setting, we do not assume any previous knowledge of
the CTRs although we do assume that such CTRs exist.
Here, we show that any mechanism that is truthful under such
a setting must follow some very rigid restrictions on its
allocation rule.
\begin{definition}[Strong Pointwise Monotonicity]
An allocation rule is said to be
{\it strongly pointwise monotone} if it satisfies: For any fixed $(b_{-i},\rho)$, if an agent $i$ with bid $b_i$ is allocated
a slot $j$ in round $t$, then $\forall \; b_i^+>b_i$, she is
allocated the same slot $j$ in round $t$.
That is if the agent $i$ receives a slot in round $t$, 
then she receives the same slot for any higher bid. For any 
lower bid, either she may receive the same slot or may loose
the impression. 
\end{definition}

\begin{theorem}
\label{thm:gen}
\noindent Let $(A,P)$ be a deterministic, non-degenerate
mechanism for the MAB, multi-slot sponsored search auction,
with unconstrained and unknown $\mu_{ij}$. Then, mechanism
$(A,P)$ is DSIC \emph{iff} $A$ is \emph{strongly pointwise monotone}
and \emph{weakly separated}. Further, the payment scheme
is given by,
$$P_i(b_i,b_{-i};\rho)=b_iC_i(b_i,b_{-i};\rho)-\int_0^{b_i}C_i(x,b_{-i};\rho)dx.$$
\end{theorem}
\beginproof
The proof is organized as follows. In step 1, we show the necessity of
the payment structure. In step 2, we show
the necessity of strong pointwise monotonicity. Step 3 proves the necessity
of weakly separatedness. Finally in step 4, we prove that
the above payment scheme in conjunction with strong pointwise  monotonicity
and weakly separatedness imply that the mechanism is DSIC.\\
\vspace*{3pt}
\noindent\underline{Step 1:} The utility structure for each agent
$i \in  N$ is \\
$U_i(v_i,(b_i,b_{-i}),\rho)=v_iC_i((b_i,b_{-i}),\rho)-P_i((b_i,b_{-i}),\rho)$

The mechanism is DSIC {\it iff} it is the best response for each
agent to bid truthfully. That is, by bidding truthfully, each
agent's utility is maximized. Thus\\
$(A,P) \mbox{ is DSIC } \Leftrightarrow \frac{dU_i}{db_i}\vert_{b_i=v_i}=0$
and $\frac{d^2U_i}{db_i^2}\vert_{b_i=v_i}\leq 0 \;\forall v_i$.

From the first order equation, we obtain,
$$b_i\frac{dC_i}{db_i}-\frac{dP_i}{db_i}=0 \; \forall b_i$$
We need $P_i(0)=0$ for normalization. Integrating the above and by
second order conditions, we need  $\frac{dC_i}{db_i}\geq 0$,
which is the clickwise monotonicity condition.

\noindent Thus, for $(A,P)$ to be DSIC, we need
\begin{eqnarray}
P_i(b_i,b_{-i};\rho)=b_iC_i(b_i,b_{-i};\rho)-\int_0^{b_i}C_i(x,b_{-i};\rho)dx \nonumber\\
  \mbox{ and } \frac{dC_i}{db_i}\geq 0 \;\forall ((b_i,b_{-i}),\rho)
\label{eqn:thm1}
\end{eqnarray}
\vspace*{3pt}
\noindent\underline{Step 2:} We first prove the necessity of
strong pointwise monotonicity by contradiction. We have seen from
(\ref{eqn:thm1}) that
$\frac{dC_i}{db_i}\geq 0 \; \forall ((b_i,b_{-i}),\rho)$
is necessary for  DSIC of $A$. We show that
if $A$ is not strongly pointwise monotone, then there exists some
allocation and realization $\rho$ for which $\frac{dC_i}{db_i}< 0$.
If $A$ is not strongly pointwise monotone, there exists
$(b_i,b_i^+,b_{-i},\rho,t) \ni$
\begin{eqnarray}
A_{ij_1}((b_i,b_{-i}),\rho,t)=1\mbox{ and }
A_{ij_2}((b_i^+,b_{-i}),\rho,t)=1,\nonumber\\
\mbox{ where }j_1\neq j_2
\label{eqn:contra_thm1}
\end{eqnarray}
Over all such counter-examples, choose the one
with the minimum $t$. By this choice, we ensure that in this
example $\forall t'<t$, we have $A_{ij}(b_i,t')=A_{ij}(b_i^+,t')$.
The only difference occurs in round $t$.
Now, consider the game instance where
$\rho_{ij_1}(t)=1,\;\rho_{ij_2}(t)=0,\; \rho_{ij}(\tau)=0 \forall
\: \tau > t $. The occurrence of such $\rho$ has non-zero
probability. Now, under $(b_{-i} ,\rho)$,
agent $i$ has the same allocation and the same number of clicks until
round $(t-1)$ independent of whether she bids $b_i$ or $b^+_i$.
However, in round $t$ with bid $b_i$, she receives a click and
with bid $b_i^+$ she does not, implying for this case that
$\frac{dC_i}{db_i}<0$. This violates the click monotonicity
requirement. So, strong pointwise monotonicity is indeed a necessary
condition for truthful implementation of MAB mechanisms under this
setting.\\
\vspace*{3pt}
\noindent\underline{Step 3:}
We prove the necessity of the weakly separatedness condition
by contradiction. That is we assume (\ref{eqn:contr_weak_sep}).
Over all such possible counter-examples of
$b_i,b_i^+,b_{-i},\rho$, choose the one with the least $t$.
Now, either $i^*=i$ or $i^*\neq i$.\\
\underline{Case 1: ($i^*=i$)}. Consider the realization $\rho'$
differing from $\rho$ only in round $t$ in the entry
$\rho_{i^*j^*}$. That is,
$\rho'_{i^*j^*}(t) = 1-\rho_{i^*j^*}(t)$ and
$\rho'_{i'j}(t'')=\rho_{i'j}(t'')\:\forall (i',j,t'')\neq
(i^*,j^*,t)$.
We can assume, $\rho_{ij}(\tau) = 0\: \forall \: \tau > t$
as the clicks in future rounds do not affect decisions in
the current round. Since $(i^*=i,j^*)$ is not part of the
allocation in round $t$ under the original bid $b_i$, the
difference between $\rho$ and $\rho'$ is not observed by the
mechanism. However, the prices computed by the payment scheme
(\ref{eqn:thm1}) to agent $i$ differ under these two
realizations. (See \cite{BABAIOFF09} for
details on why the payments are different). \\
\vspace*{3pt}
\underline{Case 2: $i^*\neq i$}. Now, choose $\rho(t)$ to be that
realization for which $(i^*,j^*)$ is strongly $i$-influential. Now,
let $t'$ be the first round $i$-influenced by $(i^*,j^*,t)$.
Consider the realization $\rho'$ which differs from $\rho$ in that
$\rho'_{i^*j^*}(t)=1-\rho_{i^*j^*}(t)$. Agent $i$'s allocation and
click information differs only in round $t'$ under the two
different realizations $\rho$ and $\rho'$. Now, let
$A_{ij_1}((b_i,b_{-i}),\rho,t')=1$ and
$A_{ij_2}((b_i,b_{-i}),\rho',t')=1$ or agent $i$ gets slot $j_1$ in
round $t'$ with bid $b_i$ under realization $\rho$ and slot $j_2$
under realization $\rho'$. Here since the two differ only in
$\rho_{i^*j^*}(t)$, by the strongly $i$-influential nature of
$(i^*,j^*,t)$ under this realization we have $j_1\neq j_2$. Without
loss of generality, let $j_1<j_2$ (or $j_1$ be the better slot,
since it is possible that one of the realizations leads to no slot
allocation). Now, we choose $\rho(t')=\rho'(t')$ in the following
manner: $\rho_{ij_1}(t')=\rho'_{ij_1}(t')=1$,
$\rho_{ij_2}(t')=\rho'_{ij_2}(t')=0$, and
$\rho_{ij}(\tau) = \rho'_{ij}(\tau) = 0, \:\forall \:\tau > t'$.
We can make such an arbitrary choice since the realization
from the round $t'$ onwards does not affect the allocation
in round $t$.

Under this choice of $\rho$ and $\rho'$, agent $i$ clearly
gets more clicks under realization $\rho$ than $\rho'$ with
bid $b_i$. Now, agent $i$'s number of clicks varies with her bid
based on only her allocation in round $t'$ which changes only
if with bid $x$ the pair $(i^*,j^*)$ is $i$-influential in round
$t$ with $t'$ earliest influenced round. With any such bid
$x$, under realization $\rho'$, agent $i$ will either get slot
$j_2$ in round $t'$ or no slot at all (by strong pointwise monotonicity),
which in turn means that she will never get a click under
realization $\rho'$ in round $t'$. Hence,
$C_i((x,b_{-i}),\rho)\geq C_i((x,b_{-i}),\rho')\forall x\leq b_i^+$.
Additionally, we have
$C_i((x_0,b_{-i}),\rho)>C_i((x_0,b_{-i}),\rho')$. Using these
relations and the non-degeneracy condition (see \cite{BABAIOFF09}
for details), we have
$P_i((b_i^+,b_{-i}),\rho)<P_i((b_i^+,b_{-i}),\rho')$.
$\rho'$ only differs from
$\rho$ in the unobserved bit $(i^*,j^*,t)$. Hence, the mechanism
fails to assign unique payment to agent $i$ leading to a
contradiction. This shows the necessity of weakly separatedness.\\
\vspace*{3pt}
\noindent\underline{Step 4:}
Finally, we show that strong pointwise monotonicity and
weakly separatedness are sufficient conditions for 
clickwise monotonicity and computability of the payments
and hence for truthfulness. Suppose $A$ is a strongly pointwise
monotone and weakly separated allocation rule.
So, it clearly satisfies the clickwise monotonicity.
Now, by the weakly separatedness condition, we
already have all the information required to calculate the
allocation of agent $i$ in every round for every bid $x<b_i$.
This is because the $i$-influential set for bids $x<b_i$
is a subset of the known influential
set, we already have all the possible click information
required for the $i$-influential sets. Additionally, by the
strong pointwise monotonicity condition, we know that for each bid
$x<b_i$ and each round $t$, either agent $i$ keeps the same
slot she had in the observed game instance $(b_i,b_{-i};\rho)$
or loses the impression altogether, that is, does not get a
click. Hence, we have all the information required to compute
$P_i(b_i,b_{-i};\rho)=b_iC_i(b_i,b_{-i};\rho)-\int_0^{b_i}C_i(x,b_{-i};\rho)dx$.
This completes the sufficiency part of the theorem.
\proofends

\subsubsection*{Implications of Strong Pointwise Monotonicity}
For a given round $t$, if an agent $i$ is allocated a slot $j$, 
then by the definition of strong pointwise monotonicity she 
receives the same slot for any higher bid that she places. If 
she lowers her bid, she may either retain the slot $j$, or lose 
the impression entirely. This leads to the strong 
restriction that an agent's bid can only decide whether or not 
she obtains an impression, and not which slot she actually gets. 
As we shall show below, this restriction has serious 
implications on the regret incurred by any truthful mechanism.

\subsubsection*{Regret Estimate}
In the single slot case it is a known result that the worst
case regret is $O(T^{2/3})$ \cite{BABAIOFF09}. So, for the
multi-slot case, the regret is $\Omega(T^{2/3})$. We show here
that the worst case regret generated in the multi-slot general
setting by a truthful mechanism is in fact  $O(T)$. 
We show this for the 2 slot, 3 agent case with an intuitive 
argument, which can be generalized.

Consider a setting with two slots and three competing 
agents, that is $m=2,\; k=3$.
Let the agents be $A_1$, $A_2$ and $A_3$. 
By Theorem \ref{thm:gen}, any truthful mechanism has to be
strongly pointwise monotone. That is, in any round, 
the bids of the agents only determine which agents will be 
displayed and not the slots they obtain.

Suppose, $A_3$'s bid $b_3<\min(b_1,b_2)$ in addition to having 
low CTRs. In this case, any mechanism that grants $A_3$ an 
impression $O(T)$ times, will have regret $O(T)$. 

So, we can assume that $A_3'$s ad gets an impression for a very
small number of times when compared with $T$. Thus, ads by $A_1$ 
and $A_2$ will appear $O(T)$  times. In each round, $A_1$ will get
either slot 1 or slot 2 independent of her bid, while the other
slot is assigned to $A_2$.

In any strongly pointwise monotone mechanism, either $A_1$ is assigned a slot 1 $O(T)$ times or slot 2. 

Without loss of generality, we assume that $A_1$ is assigned slot 1 $O(T)$ times. So, the allocation (slot 1, slot 2)$\leftrightarrow$($A_1$,$A_2$) is made $O(T)$ times.
Consider a game instance where this is not the welfare maximizing assignment, that is, the relation $(\mu_{11}b_1+\mu_{22}b_2)<(\mu_{12}b_1+\mu_{21}b_2)$ holds true. Since the slot allocation does not depend on the individual bids, such an instance can occur. In such a setting ($A_2,A_1$) would have been optimal assignment. As a result, each round having the allocation ($A_1,A_2$) incurs constant non-zero regret. Since such an allocation occurs $O(T)$ times, the mechanism has a worst case regret of $O(T)$. Hence any truthful mechanisms under the unrestricted CTR setting exhibit a high $O(T)$ regret.
\proofends

Since the strong monotonicity condition places such a severe
restriction on $A$ and also leads to a very high regret, in the
following sections we explore some relaxations on the assumption
that $\mu_{ij}$'s are unrelated. With such settings which
are in fact practically quite meaningful, we are able to prove  more
encouraging results.

\subsection{Higher Slot Click Precedence}
\label{ssec:general2}
This setting is similar to the general one discussed above in
that we do not assume any knowledge about the CTRs. However,
we impose a restriction on the realization $\rho$ that
it follows {\it higher slot click precedence} defined below.

\begin{definition}
A realization $\rho$ is said to follow \emph {Higher Slot Click
Precedence} if $\forall\: i\:\in K, \: \forall t = 1,2,\ldots,T$,
$$\rho_{ij_1}(t)=1 \Rightarrow \rho_{ij_2}(t)=1 \; \forall
j_2<j_1$$
\end{definition}
Higher slot click precedence implies that if an agent $i$ obtains
a click in slot $j_1$ in round $t$, then in that round, she
receives a click in any higher slot $j_2$. This assumption is
in general valid in the real world since any given user
(fixed by round $t$) who clicks on a particular ad when
it is displayed in a lower slot would definitely click on the
same ad if it was shown in a higher slot.

We show, under this setting, that {\it weak pointwise monotonicity} and
{\it weakly separatedness} are necessary conditions for
truthfulness. They are, however, not sufficient conditions. 
Clearly, strong pointwise monotonicity and weakly separatedness will 
still be sufficient conditions.
A weaker sufficient condition for truthfulness
under this setting is still elusive.

\subsubsection*{Implications of the Assumption}
\label{sssec:implications}

Observe that a slot-agent pair $(i,j)$ is influential in some
round $t$ only if changing the realization in the entry
$\rho_{i,j}(t)$ for some realization $\rho$ results in a
change in allocation in some future round. Crucial to the
influentiality is the fact that $\rho_{ij}(t)$ can change.

Now, consider the following situation: it has been observed
that in the game instance $((b_i,b_{-i}),\rho)$, we have
$\rho_{ij_1}(t) = 0$ where agent $i$ obtains slot $j_1$ in
round $t$. We are interested in the game instance
$((x,b_{-i}),\rho)$ where agent $i$ gets slot $j_2>j_1$
where $x<b_i$ and in knowing whether $(i,j_2)$ is an
influential pair in round $t$ for some influenced agent.
Now, since $\rho_{ij_1} = 0$ and $j_1<j_2$, by our defining
assumption, we conclude that $\rho_{ij_2}(t) = 0 \;\forall
x<b_i$. Hence, our mechanism knows that in all the relevant
cases, the realization in the given slot-agent pair never
changes. Hence, $(i,j_2)$ cannot be an influential pair for
any $j_2>j_1$ in round $t$. We will use this observation in
the proof of necessity characterization.

\begin{proposition}
Consider the setting in which realization $\rho$ follows
Higher Slot Click Precedence. Let $(A,P)$ be a deterministic
non-degenerate DSIC mechanism for this setting. Then the
allocation rule $A$ must be \emph{weak pointwise monotone} and
\emph{weakly separated}. Further, the payment scheme is given by,
$$P_i(b_i,b_{-i};\rho)=b_iC_i(b_i,b_{-i};\rho)-\int_0^{b_i}C_i(x,b_{-i};\rho)dx$$
\end{proposition}
\beginproof
The proof for the payment scheme is
identical to that in Theorem \ref{thm:gen}.
We prove the necessity of weak pointwise monotonicity and weakly
separatedness.

\noindent\underline{Step 1:} We first prove the necessity of
weak pointwise monotonicity, in a very similar fashion
to that of the necessity of strong pointwise monotonicity in Theorem
\ref{thm:gen}. The crucial difference is, while constructing
$\rho$, we have to ensure that it satisfies the higher order
click precedence. Suppose $A$ is truthful but not weakly pointwise
monotone, that is, $ \exists \: (b_i,b_i^+,b_{-i},\rho,t)$ and \\
$A_{ij_1}(b_i,b_{-i}),\rho,t = 1\mbox{ and }
     A_{ij_2}((b_i^+,b_{-i}),\rho,t)=1$ for some $j_1<j_2$.
Over all such examples, choose the one with the least $t$.
By this choice, we ensure that in this example, $\forall t'<t$,
we have $A_{ij}(b_i,t')=A_{ij}(b_i^+,t')$. The only difference
occurs in round $t$. Now, consider the game instance where
$\rho_{ij_1}(t)=1$ and $\rho_{ij_2}(t)=0$. Such realization has
a non-zero probability of occurrence. Now, under $(b_{-i} ,\rho),$
agent $i$ gets the same allocation and the same number of
clicks until round $(t-1)$ independent of whether she bids
$b_i$ or $b^+_i$. However, in round $t$ with bid $b_i$ she
gets a click and with bid $b_i^+$ she does not, implying
for this case that  $\frac{dC_i}{db_i} < 0$. This leads to a
contradiction.  So, weak pointwise monotonicity is a necessary
condition. If $A$ is not strongly pointwise monotone, does not violate
clickwise  monotonicity. That is, for truthful $A$, it
may possible that, $A_{ij_1}((b_i,b_{-i}),\rho,t) = 1$ and
$A_{ij_2}((b_i^+,b_{-i}),\rho,t) = 1$ where $j_2< j_1$.
Thus, for $A$ to be truthful, strong pointwise monotonicity may not
be necessary.

Next, we prove the necessity of the weakly separatedness
condition. Again, we prove this claim by contradiction. We
follow the same steps as in proof of Theorem \ref{thm:gen},
except we need to justify our choices of $\rho$, as it should
satisfy higher order click precedence property.\\
\vspace*{3pt}
\underline{Case 1: ($i^*=i$)}. Here, as in the previous proof we
choose $\rho'$ such that, $\rho'_{i^*j^*}(t)=1-\rho_{i^*j^*}(t)$
and $\forall \; (i',j,t'')\neq (i^*,j^*,t) \; \rho'_{i'j}(t'') =
\rho_{i'j}(t'').$ We need to show that this choice of $\rho$
does not contradict the higher slot click precedence. Now, from
our assumption, $(i^*,j^*)$ is an influential pair in round
$t$. From our observation in Section \ref{sssec:implications},
it follows that $\rho_{i^*j^*}(t)$ and $\rho'_{i^*j^*}(t)$ must
be able to take any value from $\{0,1\}$. It also forces that
$\forall j<j^*, \: \rho_{i^*j}(t)=1$. Since $(i^*,j^*)=(i,j^*)$
is not part of the allocation in round $t$ under the original
bid $b_i$, the difference between $\rho$ and $\rho'$ is
not observed by the mechanism. However, the payments by agent $i$
differ under these two realizations (see \cite{BABAIOFF09} for
details on why the payments are different).\\
\vspace*{3pt}
\noindent\underline{Case 2: $i^*\neq i$}.
Here, over all examples with $(i^*,j^*)$ influential
pair in round $t$ with influenced agent $i$ in the earliest
influenced round $t'$, we choose the one with minimum $x_0$.
In this case, our choice of $\rho(t)$ and $\rho'(t)$ is the
same as in Theorem \ref{thm:gen}, while our choice of
$\rho(t') = \rho'(t')$ differs. Again the choice of $\rho(t)$
and $\rho'(t)$ is a valid assumption by the influentiality
of $(i^*,j^*)$. Without loss of generality, let $j_1<j_2$
where $j_1$ and $j_2$ are defined as in {\it Theorem 3.1}.
Now, we choose $\rho(t') = \rho'(t')$ in the following manner:
$\rho_{ij}(t') = \rho'_{ij}(t')=1\;\forall j\leq j_1$ and
$\rho_{ij'}(t') = \rho'_{ij'}(t')=0\;\forall j'>j_1$.
We can make such an arbitrary choice since the realization
from the round $t$ onwards does not affect the allocation
in round $t$. Now, the rest of the arguments from the proof
of Theorem \ref{thm:gen} follow and lead to contradiction
that a mechanism can not distinguish between $\rho$ and $\rho'$,
however it needs to assign different payments under these
realizations. This shows the necessity of weakly separatedness.
\proofends

\subsection{When CTR Pre-estimates are Available}
\label{ssec:agentdependantctr}
In this setting, we assume the existence of some previous
database or pre-estimate of CTR values but {\it no restriction
on $\rho$}. That is, $\mu_{ij}=\frac{X_{ij}}{Y_{ij}}$ where
$X_{ij}$ is the number of clicks obtained by agent $i$ in
slot $j$ out of the $Y_{ij}$ times she obtained the slot $j$
over all past auctions. Here, in general, $\mu_{i1}\geq \mu_{i2}
\geq \ldots\geq \mu_{im}$. For our characterization, we assume
that each $\mathbb{\mu}_i = (\mu_{i1}, \mu_{i2},\ldots,\mu_{im})$
is known to the agent $i$ and the auctioneer.

In this setting, the auctioneer uses explorative rounds to
improve his estimate of the CTRs and updates the database. Then, he
makes use of his new knowledge of the CTRs in the exploitative
rounds. The payment scheme, however, only makes use of the old CTR
matrix. Under this scheme, we derive the conditions required for a 
mechanism to be {\emph truthful in expectation over $\mu$}, defined as follows. 
\begin{definition}
[Truthful in Expectation]
A mechanism is said to be truthful in expectation over $\mu$,
the CTR pre-estimate, if each of the agents believes that the
number of clicks she obtains is indeed $\sum_t\sum_j 
(\mu_{ij}A_{ij})$, which is the number of clicks she will obtain
if the CTR pre-estimate is perfectly accurate.
\end{definition}

\subsubsection{Fairness}
\label{sssec:fairness}

For this characterization, we need the notion of fair allocation
rules, as defined below.

\begin{definition}[Fair Allocation] Consider two game
instances $((b_i,b_{-i}),\rho)$ and $((b_i',b_{-i}),\rho)$
having the same slot-agent-round triplets, $(i',j',t')$ as
{\it strongly i-influential}. Let $(i^*,j^*,t)$ be such triplet
with the smallest $t'$ in which $i$ is influenced. Consider
the realization $\rho'$ differing from $\rho$ only in this
influential element $\rho_{i^*j^*}(t)$.
Then, the allocation rule A is said to be fair if for every such
pair of games it happens that\\

$\sum_j\mu_{ij}A_{ij}((b_i,b_{-i}),\rho,t')\geq
\sum_j\mu_{ij}A_{ij}((b_i,b_{-i}),\rho',t')\Leftrightarrow
\sum_j\mu_{ij}A_{ij}((b_i',b_{-i}),\rho,t')\geq
\sum_j\mu_{ij}A_{ij}((b_i',b_{-i}),\rho',t')$
\end{definition}

The intuition behind {\it fair allocations} is that changing the
realization {\it only} in a fixed strongly $i$-influential slot
generally changes agent $i$'s allocation in a predictable fashion
independent of her own bid, either improving her slot or worsening
it in the earliest influenced round, irrespective of the allocation
or realization in the rest of the game. For example, if agent $i$'s
chief competitor agent, $i'$, is strongly $i$-influential, then $i'$
not getting a click in the influential round will generally mean
that agent $i$ will go on to get a better slot than if agent $i'$
got a click, independent of $b_i$.

\subsubsection{Truthfulness Characterization}\footnote{Note, the characterization in this section would hold
even if $\mu_{ij}$ are arbitrary weights. However, while
using arbitrary weights, mechanism may charge some agents 
more than their actual willingness to pay. Also
regret in the revenue, that is loss in the revenue to the search engine will be trivially O($T$).}

\label{sssec:agentdepchar}
Here, the expected utility for the agent $i$, %
\begin{equation}
U_i(v_i,b,\rho)=(\sum_{t=1}^T\sum_{j=1}^m\mu_{ij}A_{ij}(b,t)v_i)-P_i(b,\mu)
\label{eqn:expt_utility}
\end{equation}
\begin{proposition}
\label{prop:ptmon} Let $(A,P)$ be a normalized
mechanism under this setting. Then, the mechanism is truthful in
expectation over $\mu$ {\it iff} $A$ is weakly pointwise monotone and the
payment rule is given by
$$P_i(b,\mu)=\sum_{t=1}^T\sum_{j=1}^m\mu_{ij}\{b_iA_{ij}(b,\mu,t)-\int_0^{b_i}A_{ij}(x,b_{-i},\mu,t)dx\}$$
and payments are computable.
\end{proposition}
\beginproof
In Step 1, we prove the necessity and sufficiency of the payment
structure. For the mechanism to be implemented, we need
to compute the payments of all the agents uniquely. That is,
$P_i$s need to be computable for all agents $i$. In Step 2,
we show weak pointwise monotonicity is equivalent to the second order
condition which is clickwise monotonicity in the context
of this paper.

\noindent\underline{Step1:}
The expected utility of an agent $i$ is given by (\ref{eqn:expt_utility}).\\
\noindent The $(A,P)$ is truthful {\it iff} $\frac{dU_i}{db_i}\vert_{b_i=v_i}=0$ \&
$\frac{d^2U_i}{db_i^2}\vert_{b_i=v_i}\leq 0$  $\forall v_i$.

From the first order condition we get, 
$$P_i(b,\mu)=\sum_{t=1}^T\sum_{j=1}^m\mu_{ij}\{b_iA_{ij}({b},{\it
\mu},t)-\int_0^{b_i}A_{ij}(x,b_{-i},{\it \mu},t)dx\}$$

From the second order condition, we need,
\begin{equation}
 \forall \: i,\;\sum_t\sum_j\mu_{ij}\frac{dA_{ij}}{db_i}\geq 0
\label{eqn:sec_ord_ctr_pre}
\end{equation}

\noindent\underline{Step 2:}
We show, (\ref{eqn:sec_ord_ctr_pre}) $\Leftrightarrow$ {\it 
weak pointwise monotonicity}.

\noindent (i) It is obvious that {\it weak pointwise monotonicity}
$\Rightarrow$ \\$\sum_t\sum_j\mu_{ij} \frac{dA_{ij}}{db_i} \geq 0$.
An increase in $b_i$ under a weakly pointwise monotone $A$
would result in a better slot allocation for agent $i$. This
in turn, would result in an increase in $\sum_j\mu_{ij}A_{ij}$
in each round.

\noindent (ii) Now we prove the converse.
Suppose $A$ is not weakly pointwise monotone.
That is, $\exists$ $i,b_i,b_i^+,b_{-i},\rho,\mu,t \ni \\
A_{ij}(b_i,b_{-i},\rho,\mu,t)=1$ and
$A_{ij'}(b_i^+,b_{-i},\rho,\mu,t) = 1$ where $j'>j$.
Consider the smallest such $t$.
Allocation in this round does not depend upon the realization of 
this round or of future rounds. 
We consider the instance of the game where $\rho_{ij}(t)=1$ and
$\rho_{ij'}(t)=0$ and $t$ is the last round. Such an instance
has a non-zero probability and for this instance,
$\sum_t\frac{d}{db_i}\sum_j\mu_{ij}A_{ij}<0$. This proves the
equivalence claim.
\proofends

Note, it is crucial that each $\mu_{ij}$ is a
previously known constant and cannot be defined as
$\mu_{ij}=X_{ij}/Y_{ij}$ based on the clicks in the
current $T$ rounds post facto. If we do so,
$X_{ij}/Y_{ij}$ can change with the allocation of agent $i$
in a particular game and hence, $\mu_{ij}$ would become a function
of $b_i$ and the mechanism would be no longer truthful.

For truthful implementation, the payments need to be
computable and computing the payments may involve the unobserved part of $\rho$. In the next theorem,
we show that weakly separatedness is necessary
and sufficient for computation of these payments. So, along with the
computation of payments and the above proposition,
we get,

\begin{theorem}
\label{thm:ctrpre-estimate}
Let $(A,P)$ be a mechanism for this stochastic
multi-round auction setting where $A$ is a non-degenerate,
deterministic and \emph{fair} allocation rule. Then, $(A,P)$ is
truthful in expectation over $\mu$ \emph{iff} $A$ is \emph{weakly pointwise
monotone} and \emph{weakly separated} and the payment scheme is given
by
$$P_i(b,\rho)=\sum_{t=1}^T\sum_{j=1}^m\mu_{ij}\{b_iA_{ij}({\it
b},\rho,t)-\int_0^{b_i}A_{ij}(x,b_{-i},\rho,t)dx\}$$
\end{theorem}
\beginproof
\footnote{The idea for our proof is similar to that in the characterization of
the single-slot case \cite{BABAIOFF09}; however, the details
are non-trivially different.
}
This setting/characterization works best
with old advertisers who have already taken part in a large number
of auctions. As we already have proved Proposition \ref{prop:ptmon},
we just need to show that weakly
separatedness is in fact a necessary and sufficient condition for
the computability of payments, that is, computability of\\
$\sum_{j=1}^m\mu_{ij}\int_0^{b_i}A_{ij}(x,b_{-i},\rho,t)dx$ for
each agent $i$.

\vspace*{3pt}
\noindent\underline{Step 1:} We first provide the proof for
the sufficiency of weakly separatedness. Suppose $A$ is weakly
separated. The mechanism observes and knows all allocations and the
observed realization for the game instance carried out with the
original bid vector $(b_i,b_{-i})$. Specifically, it knows
$N((b_i,b_{-i}),i,\rho,t)$ for all rounds $t$ and the respective
realizations in these slots. Now, in the game instance $(x,b_{-i})$
where $x\leq b_i$, by weakly separatedness, we have
$N((x,b_{-i}),i,\rho,t)\subseteq N((b_i,b_{-i}),i,\rho,t)$. This
means that the allocation in $i$-influential slots for game instance
$((x,b_{-i})t)$ is a subset of that in observed game instance
$((b_i,b_{-i}),\rho)$. So, the mechanism already knows all the click
information in the $i$-influential slots for the game instance
$((x,b_{-i}),\rho)$. Since the payment scheme is only interested in
the allocation of agent $i$, the realization in the unobserved slots
is unimportant and can be assumed arbitrarily. Thus, the mechanism has
complete information to compute $P_i((b_i,b_{-i}),\rho,t)$.\\
\vspace*{3pt}
\noindent\underline{Step 2:} Next, we prove the necessity of weakly
separatedness by contradiction. That is, we assume
(\ref{eqn:contr_weak_sep}) is true.
Consider a {\it complete} realization $\rho(t)$ in round $t$ for
which $(i^*,j^*)$ is strongly $i$-influential (such a realization
exists by our previous theorem) and construct the two complete
realizations $\rho$ and $\rho'$ from
$(\rho(1),\rho(2),\ldots,\rho(t-1),\rho(t))$ which only differ in
$\rho_{i^*j^*}(t)$.Over all choices of counter-examples
$(b_i,t,\rho(t),i^*,j^*)$, we choose the one which has the smallest
influenced round $t'$. Now, we compare the payment that the
mechanism has to make for this game instance at the end of $t'$
rounds under the two different realizations $\rho$ and $\rho'$.

Let $\varphi\in \{\rho,\rho'\}$. By the strong $i$-influence of
$(i^*,j^*,t)$, the agent $i$ gets different allocations in round
$t'$ under the different realizations $\rho$ and $\rho'$. This
implies,\\
$\sum_j\mu_{ij}A_{ij}((b_i,b_{-i}),\rho,t')\neq
\sum_j\mu_{ij}A_{ij}((b_i,b_{-i}),\rho',t')$. \\
Without loss of generality,
\begin{equation}
\sum_j\mu_{ij}A_{ij}((b_i,b_{-i}),\rho,t')>\sum_j\mu_{ij}A_{ij}((b_i,b_{-i}),\rho',t')
\label{eqn:1}
\end{equation}
(or agent $i$ gets a higher slot under realization $\rho$ than
$\rho'$).

By the {\it non-degeneracy} of $A$, there exists a finite interval
of bids about $b_i$ such that for every bid $x$ in this interval,
\begin{equation}
A_{ij}((x,b_{-i}),\varphi,t')=A_{ij}((b_i,b_{-i}),\varphi,t')\forall
j \label{eqn:2}
\end{equation}

Suppose $x'\in (0,b_i^+)$ is another bid such that the same
slot-agent-round set $(i^*,j^*,t)$ is strongly $i$-influential with
the same influenced round $t'$ for the game $((x',b_{-i}),\rho)$.
Then by the {\it fairness } of $A$,
\begin{equation}
\sum_j\mu_{ij}A_{ij}((x',b_{-i}),\rho,t')\geq
\sum_j\mu_{ij}A_{ij}((x',b_{-i}),\rho',t')
\label{eqn:3}
\end{equation}

From (\ref{eqn:1}),(\ref{eqn:2}), and (\ref{eqn:3}) and
using the fact that $t'$ the smallest
influenced round that is strongly $i$-influenced by the bit
$\rho_{i^*j^*}(t)$ which is the only differing bit between $\rho$
and $\rho'$, we can see that $\forall \: x'\in \:(0,b_i^+)$
\begin{equation}
\sum_j\mu_{ij}A_{ij}((x',b_{-i}),\rho,t')\geq
\sum_j\mu_{ij}A_{ij}((x',b_{-i}),\rho',t')
\label{eqn:4}
\end{equation}

\noindent and $\exists$ a finite interval $X$ around bid $b_i$ such
that $\forall x\:\in X$, we have,
\begin{equation}
\sum_j\mu_{ij}A_{ij}((x,b_{-i}),\rho,t')>\sum_j\mu_{ij}A_{ij}((x,b_{-i}),\rho',t')
\label{eqn:5}
\end{equation}

From equations (\ref{eqn:4}) and (\ref{eqn:5}), and
the fact that agent $i$'s allocation
is the same under both realizations $\rho$ and $\rho'$ until round
$t'$ (from smallest influenced round choice), we conclude that,\\
$\sum_{t=1}^{t'}\int_0^{b_i^+}\sum_j\mu_{ij}A_{ij}((x,b_{-i}),\rho,t)dx \; > \hspace*{1.5cm}$\\
$\hspace*{1cm} \sum_{t=1}^{t'}\int_0^{b_i^+}\sum_j\mu_{ij}A_{ij}((x,b_{-i}),\rho',t)dx $\\
Additionally, we can assume that there are no clicks after round $t'$.
As a result, we have $P_i(b_i^+,b_{-i},\rho)\neq P_i(b_i^+,b_{-i},\rho')$.
However, the mechanism cannot distinguish between the two
realizations $\rho$ and $\rho'$ as the only differing bit $\rho_{i^*j^*}(t)$
is unobserved. Hence, the mechanism fails to assign a unique
payment to agent $i$. This is a consequence of our initial
assumption (\ref{eqn:contr_weak_sep}).Thus if $A$ is not weakly
separated the payments are not computable. This completes
the proof. 
\proofends
\subsection{When CTR is Separable}
\label{ssec:separablectr}

In the previous setting we assumed that some pre-estimate on the CTR
matrix $[\mu_{ij}]$ existed. In real world applications, however, it
is very often the case that the slot-dependent probabilities are
known while the agent dependent probabilities are unknown. To
leverage this fact, we make a widely accepted assumption: we assume
that the click probability due to the slot is independent of the
click probability due to the agent. That is, we assume that
$\mu_{ij}=\alpha_i\beta_j$, where $\alpha_i$ is the click
probability associated with agent $i$ and $\beta_j$ is the click
probability associated with slot $j$. We also assume that the vector
$\beta=(\beta_1,\beta_2,\ldots,\beta_m)$ is common knowledge. In
general $\beta_1\geq \beta_2\ldots \geq \beta_m$. Here, any
mechanism will use the explorative rounds to try to learn the values
of $\alpha_i$ as accurately as possible.


Let $Y_{ij}$ denote the number of times that agent $i$ obtains the
impression for slot $j$, and $X_{ij}$ denote the corresponding
number of times she obtains a click. Then, we define
$\alpha'_i=avg_j\{\frac{1}{\beta_j}\frac{X_{ij}}{Y_{ij}}\}$ and
$\mu'_{ij}=\alpha'_i\beta_j$.

In this section, we assume $\frac{d\alpha'_i}{db_i}=0$ or that
$\alpha'_i$ does not change with bid $b_i$. We are
justified in making this assumption since $\alpha'_i$ is a good
estimate of $\alpha_i$ which is independent of which slot agent $i$
obtains how many times. By changing her bid $b_i$, agent $i$ can only
alter her allocations which should not predictably or significantly
affect $\alpha'$. It is
trivial to see that $\frac{d\alpha'_i}{db_i}=0\Rightarrow
\frac{d\mu'_{ij}}{db_i}$.

We model truthfulness based on the utility gained by each agent in
expectation over this $\mu'_{ij}$. That is, utility to an agent
$i$ is given by equation (\ref{eqn:expt_utility}), with $\mu$ being replaced by $\mu'$.
With the above setup, it can easily be seen that truthfulness
mechanisms under this setting have the same characterization as the
truthful mechanisms with a pre-estimate of CTR. 
\begin{theorem}
\label{thm:sepctr}
Let $(A,P)$ be a mechanism for the stochastic multi-round
auction setting where $A$ is a non-degenerate, deterministic and
\emph{fair} allocation rule. Then, $(A,P)$ is truthful in expectation
over $\mu'$ \emph{iff} $A$ is \emph{weakly pointwise monotone} and \emph{
weakly separated} and the payment scheme is given by
\begin{equation}
P_i(b,\rho)=\sum_{t=1}^T\sum_{j=1}^m\mu'_{ij}\{b_iA_{ij}({\it
b},\rho,t)-\int_0^{b_i}A_{ij}(x,b_{-i},\rho,t)dx\} \nonumber
\end{equation}
\end{theorem}
\beginproof
This theorem can be proven using similar arguments as used in the proof of Theorem \ref{thm:ctrpre-estimate}, with $\mu$
being replaced by $\mu'$.
\proofends


\section{Experimental Analysis}
\label{sec:experiments}

Since the single slot setting is a special case of the multi-slot setting, we obtain $\Omega(T^{2/3})$ as a lower bound
for the regret incurred by a truthful multi-slot sponsored search mechanism.

We have characterized 
truthful MAB mechanisms in various settings in the previous 
section. However, we have not studied MAB mechanisms in 
multi-slot auctions for regret estimation in such mechanisms
(except the $O(T)$ worst case bound we showed for the 
unconstrained case in Section 3.1). In this section, we 
present a brief experimental study on the regret of an 
truthful MAB mechanism for multi-slot sponsored search 
auction under separable CTR case.

For our study, we have picked
a simple mechanism belonging to the separable CTR case.
In the simulation, we displayed the agents in the
available slots in a round robin fashion
for the first $T^{\frac{2}{3}}$ rounds.
Then, we used the observed information on the clicks
to estimate the $\mu_{ij}$ values. The payments
were computed as per Theorem (\ref{thm:sepctr}).

\begin{figure}[!htb]
\begin{minipage}[b]{1.0\linewidth}
\centering
\includegraphics[width=\columnwidth]{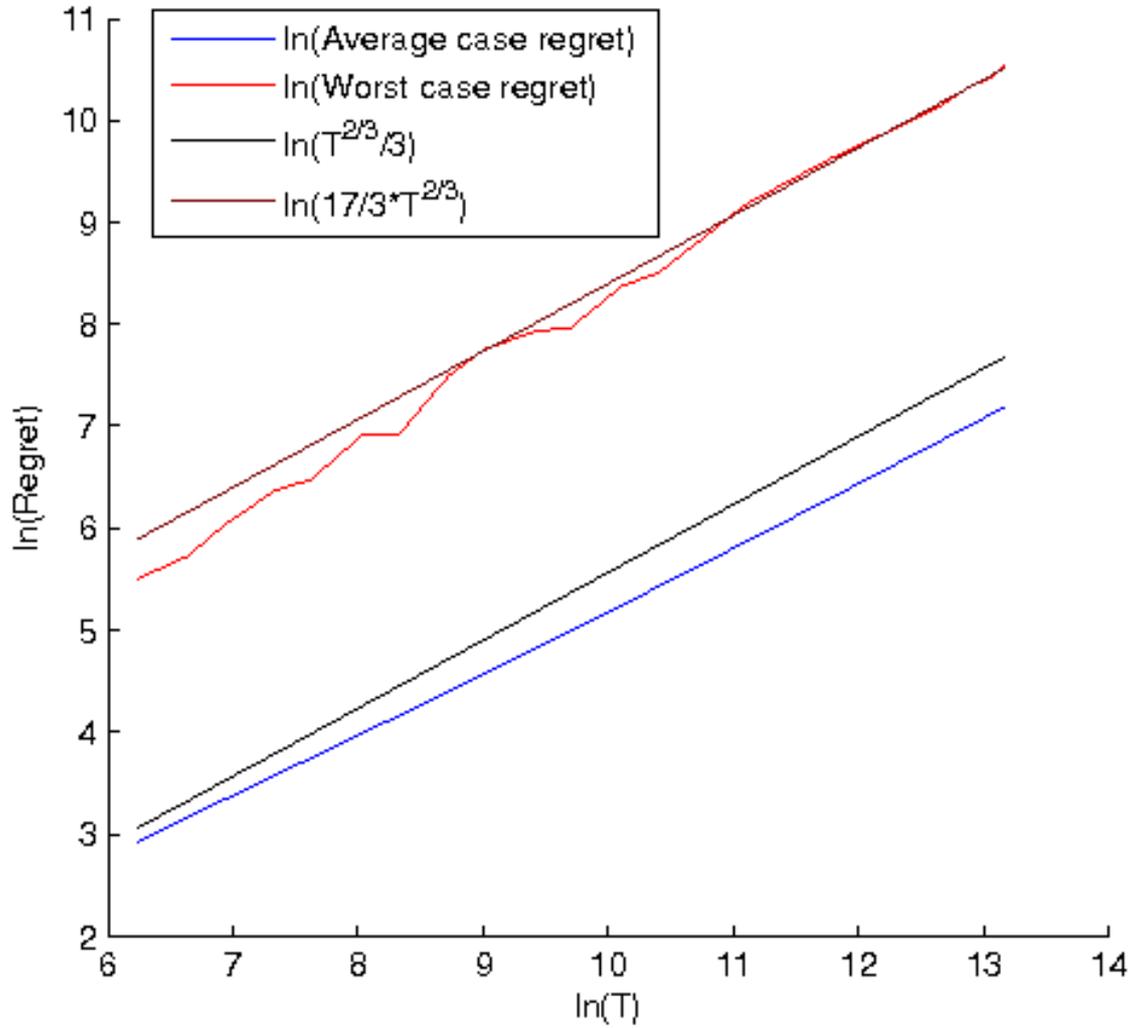}
\caption{Average Case Regret and Worst Case Regret in a Logarithmic Scale}
\label{fig:regret1}
\end{minipage}
\end{figure}

We performed simulations for various $T$ values with $k=4$ and
$m=2$. For a fixed $T$, we generated $100T$ different instances.
and estimated the average case as well as worst case regrets.
In each instance, we generate CTRs and bids randomly. 
Figure \ref{fig:regret1} depicts
$ln$(worst case regret) and  $ln$(average case regret). It is observed
that $ln$(worst case regret) is closely approximated by
$ln (\frac{17}{3} T^{2/3})$  while
$ln$(average case regret) is closely approximated by
$ln (\frac{1}{3} T^{2/3})$, clearly showing that
the worst case regret is $O(T^{2/3})$ and
the average case regret is upper bounded by $O(T^{2/3})$.


\section{Conclusion}
\label{sec:con}
In this paper, we have provided characterizations for
truthful multi-armed bandit mechanisms for various settings 
in the context of
multi-slot pay-per-click auctions, thus generalizing the
work of \cite{BABAIOFF09,DEVABUR09a} in a non-trivial way. 
The first result we proved is a negative
result which states that under the setting of unrestricted 
CTRs, any strategyproof allocation rule is necessarily 
strongly pointwise monotone. We also showed that every 
strategyproof mechanism in unrestricted CTR setting will have 
O($T$) regret. By weakening the notion of unrestricted 
CTRs, we were able to derive a larger class of strategyproof 
allocation rules.  Our results are summarized in the 
Table \ref{tab:results}.

In the auctions that we have considered, the
auctioneer cannot vary the number of slots he wishes to display. 
One possible extension of this work could be in this 
direction, that is, the auctioneer can dynamically decide 
the number of slots for advertisements.
We assume that the bidders bid their maximum willingness to pay
at the start of the first round and they would not change their 
bids till $T$ rounds.  Another possible extension would be to 
allow the agents to bid before every round. 
We are also exploring the cases where the 
bidders have budget constraints.

\bibliographystyle{plain}
\bibliography{mab}

\end{document}